%% file: sample-sigconf.tex
\documentclass[sigconf]{acmart}

\AtBeginDocument{%
  }

\setcopyright{acmcopyright}
\copyrightyear{2022}
\acmYear{2022}
\acmDOI{XXXXXXX.XXXXXXX}

\acmConference[KDD Workshop]{1st Workshop on End-End Customer Journey Optimization}{Aug 15,
  2022}{Washington DC, USA}
\acmPrice{15.00}
\acmISBN{978-1-4503-XXXX-X/18/06}
\usepackage{algorithm}
\usepackage[noend]{algpseudocode}
\usepackage{amsmath}

\begin{document}

\title{Fine-Grained Session Recommendations in E-commerce using Deep Reinforcement Learning}

\author{Diddigi Raghu Ram Bharadwaj }
\affiliation{%
  \institution{Myntra Designs Pvt. Ltd.}
  \country{India}}
\email{diddigi.bharadwaj@myntra.com}

\author{Lakshya Kumar}
\affiliation{%
  \institution{Myntra Designs Pvt. Ltd.}
  \country{India}}
\email{lakshya.kumar@myntra.com}

\author{Saif Jawaid}
\affiliation{%
  \institution{Myntra Designs Pvt. Ltd.}
  \country{India}}
\email{saif.jawaid@myntra.com}

\author{Sreekanth Vempati}
\affiliation{%
  \institution{Myntra Designs Pvt. Ltd.}
  \country{India}}
\email{sreekanth.vempati@myntra.com}

\renewcommand{\shortauthors}{Diddigi et al.}

\begin{abstract}
 Sustaining users' interest and keeping them engaged in the platform is very important for the success of an e-commerce business. A session encompasses different activities of a user between logging into the platform and logging out or making a purchase. User activities in a session can be classified into two groups: Known Intent and Unknown intent. Known intent activity pertains to the session where the intent of a user to browse/purchase a specific product can be easily captured. Whereas in unknown intent activity, the intent of the user is not known. For example, consider the scenario where a user enters the session to casually browse the products over the platform, similar to the window shopping experience in the offline setting. While recommending similar products is essential in the former, accurately understanding the intent and recommending interesting products is essential in the latter setting in order to retain a user. In this work, we focus primarily on the unknown intent setting where our objective is to recommend a sequence of products to a user in a session to sustain their interest, keep them engaged and possibly drive them towards purchase. We formulate this problem in the framework of the Markov Decision Process (MDP), a popular mathematical framework for sequential decision making and solve it using Deep Reinforcement Learning (DRL) techniques. However, training the next product recommendation is difficult in the RL paradigm due to large variance in browse/purchase behavior of the users. Therefore, we break the problem down into predicting various product attributes, where a pattern/trend can be identified and exploited to build accurate models. We show that the DRL agent provides better performance compared to a greedy strategy. 
\end{abstract}


\maketitle

\input{introduction}
\input{relatedwork}
\input{model}

\input{results}
\input{conclusions}

\bibliographystyle{ACM-Reference-Format}
\bibliography{sample-sigconf}

\end{document}

%% file: introduction.tex
\section{Introduction}
\begin{figure}
  \includegraphics[scale = 0.27]{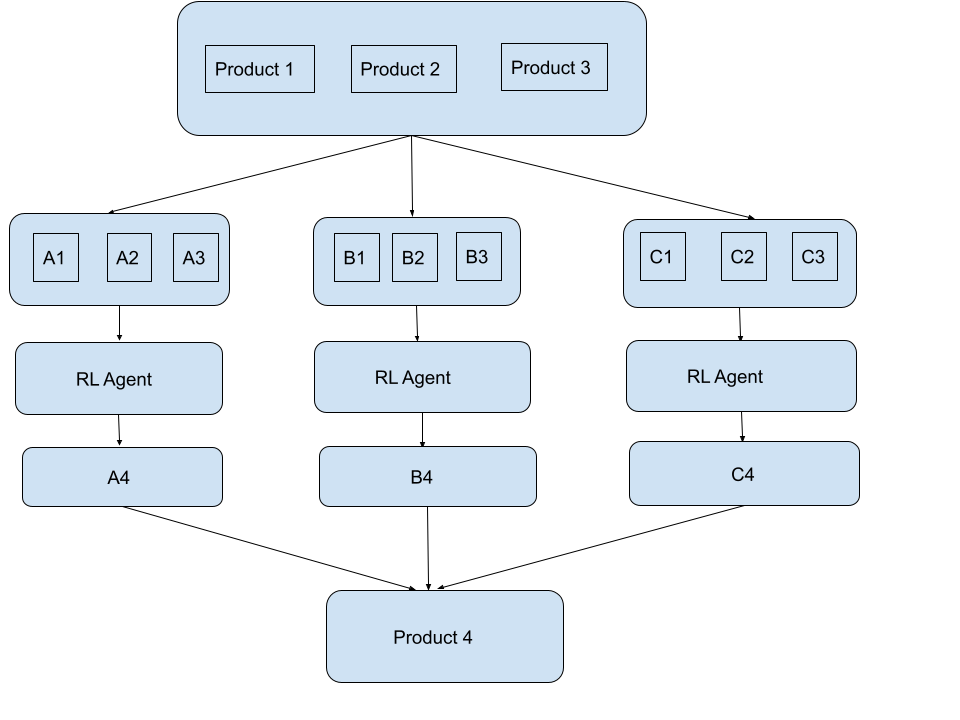}
  \caption{Proposed architecture of session recommendation. The previous $k$ (three in this example) products are considered for recommending product at instant $k+1$. First, the attributes of the products ($A,B,C$) are extracted. Next, these attributes are then fed to the RL agents to generate the next attribute recommendations. Finally, the attribute recommendations are combined to obtain the product to be recommended at instant $k+1$ (fourth in this example).}
  \label{f1}
\end{figure}
An e-commerce platform provides a promising alternative to the traditional retail business where users can browse, compare, make purchases and get delivery from the comfort of their homes. This setting has become even more popular recently due to the pandemic, where contactless purchase and delivery was preferred over the traditional shopping experience. This increased digital footprint of users on the platform led to an exponential growth of customer data like browsing history, purchase history, session activity, etc. From the e-commerce business point of view, it is paramount to make optimal use of this data not only to increase the profits of the business but also to improve the customer experience by recommending appropriate items.

Session recommendation is an essential subproblem in the recommendation system where the objective is to recommend the items to the users primarily based on the user's activity thus far in the session. It differs from the traditional recommendation systems that use the user's historical browse/purchase interactions data. On the contrary, each session is treated independently in session recommendation problems. This is due to the difficulty in learning the dependencies between the different sessions of the user. Moreover, the sessions might not be correlated, and the user's prior history (before the session starts) is not always available for cold start cases. 
Although appealing due to its simplicity, this paradigm is challenging to solve due to the inherent heterogeneity in the users' activities across the platform and varying goals associated with it. To understand this, consider the following example: In the off-line shopping scenario, a sequence of items is suggested to the customers by the sales executive based on the explicit feedback provided by the customer. It is imperative that the executive suggests the items to sustain the customers' interest and ultimately drive them towards a purchase. Each item that is suggested has an immediate impact (i.e., customers might either like and continue exploring it or exit the store) and also influences the future actions of the customers (i.e., making a purchase). Emulating this in the online scenario requires capturing this dynamic nature of the problem and balancing short-term and long-term goals. This motivates us to formulate it in the framework of the Markov Decision Process (MDP) \cite{puterman2014markov,bertsekas1996neuro}. We define a session as a sequence of events of a user until it leads to one of the following (a). purchase (b). user exits the session. This work aims to optimally recommend a sequence of products that could potentially lead to a purchase event.  

Reinforcement Learning (RL) \cite{sutton1998introduction} is a popular model-free paradigm for solving an MDP problem. Here, we train an agent to make optimal decisions based only on the trajectories of the environment. When the number of states and actions in the environment is very high, one resorts to function approximation architectures. RL algorithms combined with neural network architectures, i.e., Deep RL, have achieved a lot of success in recent times \cite{mnih2013playing,mnih2015human}. In this work, we train a Deep RL agent to recommend a sequence of products in a session. It is important to note that the traditional RL setup where agents learn by exploring different actions is not a favorable setting for our problem due to a large number of products in the e-commerce space. Hence, we train the algorithm under an off-policy setting using the users' historical session data. The dataset considered in this work is compiled from the click-stream data of users on the Myntra e-commerce platform, one of the largest fashion e-commerce in India. 

Training a deep RL agent to recommend the products directly is not practical due to heterogeneity (in terms of different attributes) in the users' browsing history. There are two problems associated with this training paradigm. First, the number of products is huge (which constitutes the action space). Second, there might not be a definite trend that can be learned from this data, making the training very unstable. For example, consider a scenario where two users browse products in a sequence that is similar in every regard (like product type, color) except a specific attribute `brand.' Say the session of the former user ends up in a purchase, whereas the latter is a non-purchase session. The effectiveness of RL training lies in the generalization of actions and hence the sessions of such nature will lead to unstable learning. To mitigate this problem, we propose a divide-and-conquer approach where multiple RL agents will be trained to recommend various attributes of the products, and these recommendations will be combined at the end to generate product recommendations. This is illustrated in Figure \ref{f1}. First, attributes of the products (like color, product type, and brand) are extracted. These attributes are then sent as inputs to independent DRL models to obtain the recommendations for attributes in the next time instant. Finally, the products that match the attributes are presented as final recommendations to the user. 

The overall contributions of the paper are as follows:
\begin{itemize}
    \item We mathematically formulate the problem of User Session recommendation in the framework of MDP.
    \item We propose a Deep Q-Learning based model to predict the next products within a user session while optimizing for purchase intent.
    \item We compare the proposed model with a similarity-based baseline model to showcase our proposed approach's efficacy. 
\end{itemize}

%% file: relatedwork.tex
\section{Related Work}
Recommendation systems deal with building algorithms for recommending products to the user to meet various objectives like user personalization, increased engagement rate, and improving business goals. The idea here is to accurately predict users' interest and recommend products that meet their expectations. Recommendation systems finds its applications in various domains like news recommendation \cite{liu2010personalized,li2019survey,zihayat2019utility,lian2018towards,zheng2018drn}, movie recommendation \cite{subramaniyaswamy2017personalised,reddy2019content,wu2018movie,ahmed2018movie} etc. The importance of recommendation systems is even more pronounced in the e-commerce business, where the buying and selling of products are performed virtually online. Therefore, it is imperative from the business point of view to recommend relevant and specific products to the users to sustain their interest over a long period. As a result, a lot of research has been dedicated to build good recommendation systems \cite{wei2016survey} in recent times to solve problems like Click-Through-Rate (CTR) prediction \cite{zhou2018deep,zhou2019deep,feng2019deep,huang2019fibinet}, intent and purchase prediction \cite{huang2019online,dou2020online,yeo2017predicting} etc. 

Deep Learning is a popular class of machine learning algorithms that uses artificial neural networks to learn and derive required patterns from the input data. We will now discuss some popular deep learning techniques proposed in the literature for solving the recommendation problem. In \cite{wei2017collaborative}, two deep learning algorithms based on the collaborative filtering technique have been proposed to handle cold-start problems. In \cite{fu2018novel}, a deep learning model has been deployed to simulate the interaction between item and user by feeding the pre-trained representations of item and user as input to the model. In \cite{liu2016recurrent}, Recurrent Neural Networks (RNNs) have been deployed to generate recommendations from user reviews. 

However, these techniques do not efficiently capture the dynamic nature of the recommendation problem. Reinforcement Learning (RL) models have the capability to handle the dynamic nature by maximizing the expected long-run objective and hence is the right paradigm for this problem. In \cite{xin2020self}, a self-supervised RL algorithm, where the output of the RL has been used as a regularizer for self supervised learning, has been proposed. In \cite{munemasa2018deep}, a recommender system based on a deep RL algorithm has been proposed. In \cite{afsar2021reinforcement,chen2021survey}, a detailed survey on the application of RL to the recommendation systems along with potential future directions has been provided. 

We now discuss the literature on the session recommendation problem. 
Session recommendation models have received a lot of attention in recent years. In \cite{hidasi2015session,ruocco2017inter}, RNN based models are proposed to solve the session recommendation. In \cite{smirnova2017contextual}, context information is included in the modeling of RNN and proposes a new class of Contextual RNNs. 
The RNN-based model is further improved by \cite{hidasi2018recurrent} where a novel loss function is used that results in improved performance. \cite{li2017neural} studies the attention mechanism that selects the most salient information for session recommendation and proposes Neural Attentive Recommendation Machine (NARM) to solve this problem. In \cite{bogina2017incorporating}, the authors have studied the impact of incorporating dwell times in the structure of RNNs and showed improved performance over standard RNN-based recommendations. In \cite{quadrana2017personalizing}, hierarchial RNNs have been proposed to improve the quality of recommendations. 

RL has also been used for session recommendation in \cite{shani2005mdp,taghipour2008hybrid}. However, they do not make use of function approximation networks leading to state and action space explosions. In \cite{zhao2017deep}, Deep RL techniques have been utilized to solve the session recommendation problem. However, these do not exploit the recurrent nature between the states in the session. In our work, we use Deep Recurrent Q-Network (DRQN) \cite{hausknecht2015deep} to tackle the session recommendation problem. As our approach does not require prior history or user information, it can also be used in cold-start recommendation problems. Moreover, as the agent is trained at the attribute-level, the recommendations can be analysed to understand the user's affinities towards these attributes. The closest work to ours is \cite{zou2019reinforcement} where an RL algorithm has been proposed to optimize for the long-term engagement in feed streaming recommendations. Our work differs from \cite{zou2019reinforcement} in the following ways:
\begin{itemize}
    \item \cite{zou2019reinforcement} considers the problem of a sequence of recommendations in feed streaming data. In contrast, our model deals with recommending products (that are computed as a function of attributes) in a session. 
    \item The reward structure in \cite{zou2019reinforcement} is aimed at improving the long-term engagement while our reward function optimizes for a purchase. 
    \item The information about the user is also part of the state space in the MDP proposed by\cite{zou2019reinforcement} whereas our models treat each session independently and do not take into consideration the user data.  
\end{itemize}

%% file: model.tex
\section{Model}
We model the session recommendation problem in the framework of Markov Decision Process (MDP). An MDP is characterised by the tuple $<\mathbf{S,A,P,R,\gamma}>$. Here $\mathbf{S}$ denotes the state space, $\mathbf{A}$ denotes the action space, $\mathbf{P}$ denotes the probability transition matrix, $\mathbf{R}$ is the single-stage reward function and $0 \leq \mathbf{\gamma} < 1$ is the discount factor. 

In the following, we describe the MDP model for recommending a sequence of a attributes. 
\begin{itemize}
    \item \textbf{State Space:} This constitutes the necessary and sufficient information to make a decision. In our problem of recommending a sequence of attributes, the information of the previous $k$ attributes browsed by the user becomes the state space. This information of the attributes is captured in the form of embeddings, which are obtained by training language models like Word2Vec \cite{mikolov2013efficient}. 
    \item \textbf{Action Space:} This constitutes the possible actions that can be taken at each step in a session. In our problem, the set of all attributes is the action space.
    \item \textbf{Reward Function}: Based on the action $a$ chosen in a given state $s$, we obtain an immediate reward $R(s,a)$ from the environment. In our problem formulation, we formulate the reward structure as follows
    \begin{enumerate}
        \item At every intermediate step of a session, the normalized dwell time (i.e., time spent by the user browsing the product) is considered to be the reward.
        \item If the session has ended in a purchase, the reward is set to +10.
        \item If the session has not ended in a purchase, the reward is set to 0. 
    \end{enumerate}
This reward structure motivates the RL agent to recommend attributes that maximize not only the user's immediate interest (captured in the form of the dwell time) but also those that ultimately lead to a purchase. An important point to note here is that the rewards are not stationary, i.e., $R(s,a)$ for a fixed state, and action $(s,a)$ changes over time. Therefore, we treat the reward as a random variable, which, as we explain later, is efficiently handed in the RL paradigm. 

\item \textbf{Probability Transition Matrix}: Based on the action $a$ chosen in a given state $s$, the system transitions to a new state which is sampled from the distribution $P(.|s,a)$. In our problem, the next state is the next $k$ attributes on the rolling window basis. 
\end{itemize}

The objective of the RL agent is to find an optimal policy $\pi^*: S \xrightarrow[]{} A$, which is a mapping from state space to action space. It represents the optimal action that needs to be taken in any given state. To compute this optimal policy $\pi^*$, the first step is to compute the Q-value function that satisfies the following Bellman equation:
\begin{align}\label{e1}
    Q(s,a) = E\Big[R(s,a) +\gamma \displaystyle \max_{b} Q(s',b)\Big], ~ \forall (s,a) \in (S \times A),
\end{align}
where $E[.]$ is the expectation over the next state $s'$ and reward function $r$. Intuitively, Q(s,a) specifies the score (combination of immediate and future rewards) that is obtained when action $a$ is taken in state $s$. Hence, the optimal policy can be obtained from Q-value as follows:
\begin{align}\label{e2}
    \pi^*(s) = \arg \displaystyle \max_{a} Q(s,a), ~ \forall s \in S.
\end{align}

\begin{figure}
  \includegraphics[scale = 0.3]{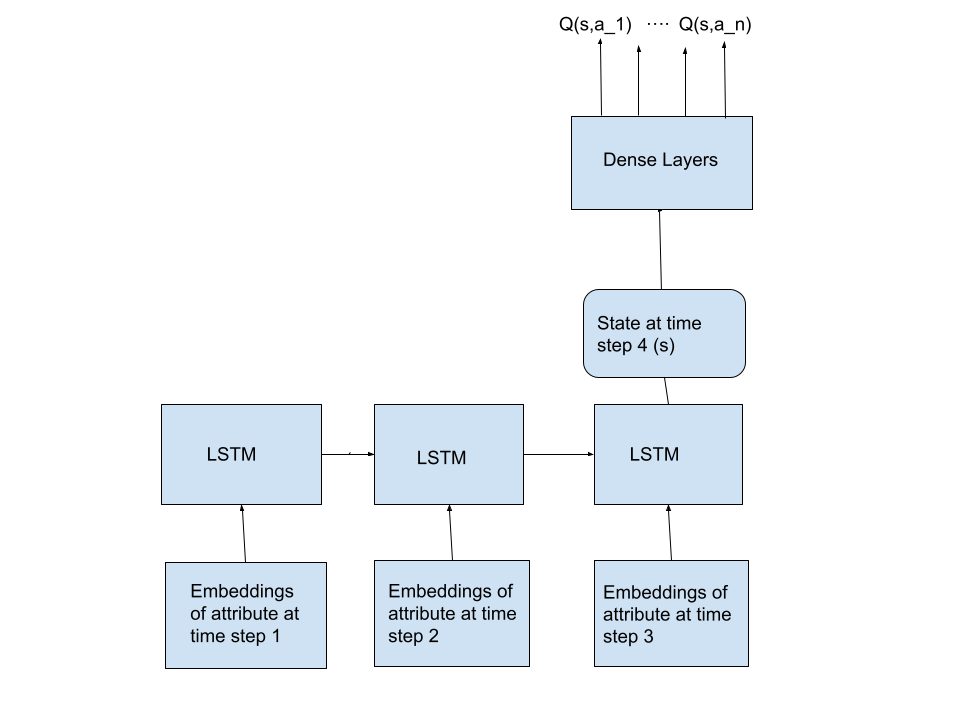}
  \caption{A single instance of the proposed algorithm. The embeddings of $k$ (three in this example) previous attributes are fed to the LSTM network to obtain a rich representation of state. This state is passed through dense layers to obtain Q-values corresponding to all actions.}
  \label{f2}
\end{figure}

In order to solve for \eqref{e1} and obtain the optimal policy as in \eqref{e2}, we make use of the popular RL algorithm Deep Q-learning Network (DQN) that makes use of neural networks. This algorithm can handle continuous state spaces and hence is a right fit for our problem. In DQN, the $Q-$values are approximated as follows:
\begin{align}
    Q(s,a) \approx f(s,a,\theta), \forall (s,a) \in (S \times A),
\end{align}
where $f$ is a non-linear function of state $s$, $a$ and $\theta$ is the parameter of the neural network. These parameters are trained by minimizing the loss function at each iteration $i$, given by \cite{mnih2013playing}:
\begin{align}\label{e3}
    \mathcal{L}_i(\theta_i) = \displaystyle E_{(s,a) \sim \rho(.)} \Big[(y_i - f(s,a,\theta_i)^2  \Big],
\end{align}
where $\rho(.)$ is the distribution from where states and actions are sampled and
\begin{align}
    y_i = \displaystyle E_{s' \sim P(.|s,a)} \Big[R(s,a) + \gamma \displaystyle \max_{b}f(s',b,\theta_{i-1}) \Big].
\end{align}
Optimisation algorithms like stochastic gradient descent could be applied to iteratively improve the parameters $\theta$ of the neural network. Here, the gradient function is obtained by differentiating \eqref{e3} with respect to parameters $\theta$ given by \cite{mnih2013playing}:
\begin{align}
    \nabla_{\theta_i} \mathcal{L}_i(\theta_i) = \displaystyle E_{(s,a) \sim \rho(.), s' \sim P(.|s,a)} \Big[&\Big(R(s,a) + \gamma \displaystyle \max_{b}f(s',b,\theta_{i-1}) \nonumber \\ &- f(s,a,\theta_i)\Big) \nabla_{\theta_i}f(s,a,\theta_i) \Big].
\end{align}
As described above, the embeddings of previous $k$ attributes constitute the state which is fed as input to the RL algorithm. The following are some of the possible options to obtain the input state:
\begin{itemize}
    \item Concatenation of all $k$ previous embeddings. In this representation, the input scales quadratically with $k$, hence not an optimal choice.
    \item Average of $k$ embeddings. Here, all the $k$ embeddings are given equal weight and hence might not be an optimal choice in practical settings. 
    \item Weighted average of $k$ embeddings. This enables one to give different weights to different embeddings, for example, higher weight to recent attributes. 
\end{itemize}
Even though the weighted average of $k$ embeddings input representation looks appealing, it fails to capture the dependence among the attributes. The relation between the attributes might form a crucial form of the state space that aids in optimal decision-making.
While optimising for long-term reward, it is very important to capture the sequential relationship and hence we rely on LSTM rather than average and weighted average. 
This motivates us to make use of the architecture shown in Figure \ref{f2}, similar to the Deep Recurrent Q-Network (DRQN). The pseudo-code of our proposed algorithm is described in Algorithm \ref{alg}. The algorithm is trained on the session data where each session is of form $s_i = \{a_n,d_n\}_{n=0}^{N_i}$, where $a_n$ is the attribute of the product that was browsed by the user, $d_n$ is the dwell time, at instant $n$ and $N_i$ is the number of steps in the session $i$. In each step of a session, the embeddings of the previous $k$ attributes are processed through an LSTM to obtain a representation of the state. Subsequently, the state is passed through a series of dense layers to obtain $Q-$ values. Finally, the parameters of the neural network are updated by performing stochastic gradient descent on loss function $\mathcal{L}(\theta),$ which is shown in eq.\eqref{e3}. 
\begin{algorithm}[h!]
\caption{DRQN- Recommendation}\label{alg}
\begin{algorithmic}[1]
\State \textbf{Input:} Session Data $\mathcal{S}= \{s_i\}_{i=0}^{\infty}$, where each session $s_i = \{a_n,d_n\}_{n=0}^{N_i}$. 
\State \textbf{Input:} k $\xleftarrow[]{}$ Number of previous steps considered as state. 
\State \textbf{Input:} Function $E():$ Maps attributes to embeddings. 
\State \textbf{Input:} $\alpha \xleftarrow{}$ Step-size.
\State Initialize the parameters of the neural network $\theta$.
\For{$i = 0,\ldots,\infty$}
\For{$n = k,\ldots,N_{i}$}
\State State $s = LSTM(E(a_{n-k}),E(a_{n-k+1}),\ldots,E(a_{n-1}))$
\State Action $a = a_n$.
\State Reward $r = d_n$
\State Next State $s' = (E(a_{n-k+1}),\ldots, E(a_{n}))$
\If{$n == N_i$}{ .\[
    y_n= 
\begin{cases}
    10,& \text{if session ended in a purchase} \\
    0,              & \text{otherwise}
\end{cases}
\]}
\Else{ $y_n = r +\gamma \displaystyle \max_{b} f(s',b,\theta)$}.
\EndIf
\State $\theta \xleftarrow[]{} \theta + \alpha \nabla_{\theta}\mathcal{L}_n(\theta) $

\EndFor
\EndFor
\end{algorithmic}
\end{algorithm}

%% file: results.tex
\section{Experiments and Results}
In this section, we discuss the experiments and results of the proposed algorithm. We train our proposed algorithm to recommend a single attribute in the sequence. 
The session data for the experiments is constructed from a large corpus of click-stream data of the users collected from the Myntra logs. Dataset and experiment related details are mentioned in Table \ref{ds-details}.
\begin{table}[]
\begin{tabular}{|l|l|}
\hline
The number of training sessions   & 33924 \\ \hline
The number of testing sessions    & 2809  \\ \hline
The value of ``k''                & 4     \\ \hline
The value of ``$\gamma$''                & 1     \\ \hline
\end{tabular}
\caption{Dataset and experimental details}
\label{ds-details}
\end{table}

The LSTM is constructed with 64 hidden states through which the state representation is obtained. The state is then passed through two dense layers with 64 hidden neurons in each layer. We refer to our proposed algorithm as ``DRQN-Recommendation''. 

For comparison purposes, we also propose an algorithm ``Similar Attributes''. In this algorithm, at every step in the session, the attribute that is similar (w.r.t cosine similarity) to the weighted average of previous $k$ attributes (where $k^{th}$ attribute is given more weight compared to $(k-1)^{st}$ and so on) is recommended. It is important to note that, although this algorithm is myopic in nature, it performs well in many practical settings as the users tend to prefer similar attributes in a given session. 

The comparison metric considered is ``Hit rate $@$ 10''. It is calculated as follows. In each step $t$ of the test session, we first obtain a list of the top $10$ attributes recommended by the algorithm. If the list contains the actual attribute browsed by the user at time $t$, then it is considered a hit (a value of $1$ is assigned). Else, it is considered as no-hit (value of $0$ is assigned). These values are averaged across all steps of all the test sessions and compiled as ``Hit rate $@$ 10''. 
\begin{table}[]
\begin{tabular}{|c|c|}
\hline
\textbf{Algorithm}           & \textbf{Hit Rate $@$ 10} \\ \hline
\textbf{DRQN-Recommendation} & 0.6736                   \\ \hline
\textbf{Similar Attributes}  & 0.5334                   \\ \hline
\end{tabular}
\caption{Performance of proposed algorithms}
\label{t1}
\end{table}
In Table \ref{t1}, we present the results of our experiments. We can observe that the ``DRQN-Recommendation'' has a higher hit rate than the ``Similar Attributes''. Please note that the current DRL model is trained to recommend a single attribute. Training and deploying multiple models (one for each attribute) and combing the results to generate a product recommendation will be part of our future work. 

%% file: conclusions.tex
\section{Conclusions and Future Work}
In this work, we have proposed a novel and efficient architecture for next product recommendations in a session. Our architecture utilizes a divide-and-conquer approach where the items' attributes are recommended, which are then combined to generate product recommendations. The objective here is to recommend products to the user in a session to sustain their interest and drive them towards potential purchases. 
We formulated the problem in the framework of MDP and utilized the Deep recurrent RL algorithm to obtain optimal policies. 

In the future, we would like to extend the training of the DRL algorithm to recommend multiple attributes and develop an end-to-end model. Also, we would like to develop other novel comparison metrics that can efficiently capture the behavior of the RL agent. performing A/B test and observe the metrics like average purchase, RPU (Revenue Per User), Average CTR etc is also a part of our future work. Moreover, applying off-policy Actor-Critic algorithms \cite{haarnoja2018soft} for this problem would be an interesting future direction.